\documentclass[11pt]{article}

\usepackage[body={6in,8.8in}]{geometry}

\usepackage{txfonts}

\usepackage{graphicx}

\usepackage[authoryear,comma,longnamesfirst,sectionbib]{natbib} 

\begin{document}


\title{\textbf{Detecting epistasis via Markov bases}}
{\renewcommand{\thefootnote}{\fnsymbol{footnote}}
\author{Anna-Sapfo Malaspinas\footnote{Department of Integrative Biology, University of California at Berkeley\newline \indent \hspace{0.2cm} Email: sapfo@berkeley.edu} $$ and Caroline Uhler\footnote{Department of Statistics, University of California at Berkeley\newline \indent \hspace{0.2cm} Email: cuhler@stat.berkeley.edu \newline\newline Anna-Sapfo Malaspinas is supported by a Janggen-Poehn fellowship. Caroline Uhler is supported by an International Fulbright Science and Technology Fellowship.}}}
\date{}
\maketitle

\begin{abstract}
Rapid research progress in genotyping techniques have allowed large
genome-wide association studies. Existing methods often focus
on determining associations between single loci and a specific phenotype. However, a particular phenotype is usually the result of complex relationships between multiple loci
and the environment. In this paper, we describe a two-stage method for detecting epistasis by combining the traditionally used single-locus search with a
search for multiway interactions. Our method is
based on an extended version of Fisher's exact test. To perform this test, a Markov chain is constructed on
the space of multidimensional contingency tables using the elements of a Markov basis as moves.
We test our method on simulated data and compare it to a two-stage logistic regression method and to a fully Bayesian method, showing that we are able to detect the interacting loci when other methods fail to do so. Finally, we apply our method to a genome-wide data set consisting of 685 dogs and identify epistasis associated with canine hair length for four pairs of SNPs.
\end{abstract}

\section{Introduction}
\label{general}

Conditions such as cancer, heart disease, and diabetes, which have significant genetic components, are the most common causes of mortality in developed countries. Therefore, the mapping of genes involved in such complex diseases represents a major goal of human genetics. 
However, genetic variants associated with complex diseases are hard to detect, because they show very little effect independently and have a low penetrance. These genetic variants likely interact to produce the disease phenotype. However, so far there has been only little evidence for the presence of multilocus interaction in complex diseases. For a recent review see (\cite{Cor09}).

Recent development of methods to screen hundreds of thousands of SNPs has allowed the discovery of over 50 disease susceptibility loci with marginal effects (\cite{McCarthy}). Genome-wide association studies have hence proven to be fruitful in understanding complex multifactorial traits. The quasi-absence of reports of interacting loci, however, shows the need for better methods for detecting not only marginal effects of specific loci, but also interactions of loci. Although some progress in detecting interactions has been achieved in the last few years using simple log-linear models, these methods remain inefficient to detect interactions for large-scale data (\cite{ACA+07}).

Many models of interaction have been presented in the past, as for example the additive model or multiplicative model (\cite{Marchini}). The former model assumes that the SNPs act independently, and a single marker approach seems to perform well. In the multiplicative model, SNPs interact in the sense that the presence of two (or more) variants have a stronger effect than the sum of the effects of each single SNP. We will discuss such models in more detail in Section \ref{interaction_models}. A complete classification of two-locus interaction models has been given by  \cite{HY08}.

In the method described in this paper, we suggest first reducing the potential interacting SNPs to a small number by filtering all SNPs genome-wide with a single locus approach. The loci achieving some threshold are then further examined for interactions. Such a two-stage approach has been suggested by \cite{Marchini}. For some models of interaction, they show that the two-stage approach outperforms the single-locus search and performs at least as well as when testing for interaction within all subsets of $k$ SNPs. These results motivate taking a two-stage approach.

Single locus methods consider each SNP individually and test for association based on differences in genotypic frequencies between case and control individuals. Widely used methods for the single-locus search are the $\chi^2$ goodness-of-fit test or Fisher's exact test together with a Bonferroni correction of the p-values to account for the large number of tests performed. We suggest using Fisher's exact test as a first stage to rank the SNPs by their p-value and select a subset of SNPs, which are then further analyzed. Within this subset, it is desirable to test for interactions with an exact test. We suggest using Markov bases for this purpose.




In Section \ref{method}, we define three models of interaction and present our algorithm for detecting epistasis using Markov bases in hypothesis testing. In Section \ref{results}, we test our method on simulated data and make a comparison to logistic regression and BEAM, a Bayesian approach suggested by \cite{BEAM}. Finally, we run our algorithm on a genome-wide dataset from dogs (\cite{CNQ+09}) to test for epistasis related to canine hair length.

\section{Method}
\label{method}

\subsection{Models of interaction}
\label{interaction_models}

In this paper, we mainly study the interaction between two SNPs and a binary phenotype, as for example the disease status of an individual. However, our method can be easily generalized for studying interaction between three or more SNPs and a phenotype with three or more states. We show a generalization in Section \ref{dogs}, where we analyze a genome-wide dataset from dogs and, inter alia, test for interaction between three SNPs and a binary hair length phenotype (short hair versus long hair). 

The binary phenotype is denoted by $D$ and takes values 0 and 1. We assume that the SNPs are polymorphic with only two possible nucleotides. The two SNPs are denoted by $X$ and $Y$, taking values 0, 1, and 2 corresponding to the three possible genotypes. We investigate three different models of interaction: a control model,
an additive model, and a multiplicative model.  The parameterization is given in the following tables showing the odds of having a specific phenotype
$$\frac{\mathbb{P}(D=1 | \textrm{genotype})}{\mathbb{P}(D=0 | \textrm{genotype})}.$$
\bigskip

\begin{itemize}
\item \textbf{Control model:} $\qquad\qquad\;\;\;\;$
\begin{tabular}{l l | c c c}
&&\multicolumn{3}{c}{$\mathbf{Y}$}\\
& & 0 & 1 &  2
\\ \hline & 0 & $\epsilon$ & $\epsilon$ & $\epsilon$
\\ $\mathbf{X}$&1 & $\epsilon$ & $\epsilon$ & $\epsilon$ \\ & 2 & $\epsilon$ & $\epsilon$ & $\epsilon$
\end{tabular}
\bigskip

\item \textbf{Additive model:} $\qquad\qquad\;\;$
\begin{tabular}{l l | c c c}
&&\multicolumn{3}{c}{$\mathbf{Y}$}\\
& & 0 & 1 &  2
\\ \hline &0 & $\epsilon$ & $\epsilon\beta$ & $\epsilon\beta^2$
\\ $\mathbf{X}$&1 & $\epsilon\alpha$ & $\epsilon\alpha\beta$ & $\epsilon\alpha\beta^2$ 
\\ &2 & $\epsilon\alpha^2$ & $\epsilon\alpha^2\beta$ & $\epsilon\alpha^2\beta^2$
\end{tabular}
\bigskip

\item \textbf{Multiplicative model:} $\qquad$
\begin{tabular}{l l | c c c}
&&\multicolumn{3}{c}{$\mathbf{Y}$}\\
& & 0 & 1 &  2
\\ \hline & 0 & $\epsilon$ & $\epsilon\beta$ & $\epsilon\beta^2$
\\ $\mathbf{X}$&1 & $\epsilon\alpha$ & $\epsilon\alpha\beta\delta$ & $\epsilon\alpha\beta^2\delta^2$ \\ &2 & $\epsilon\alpha^2$ & $\epsilon\alpha^2\beta\delta^2$ & $\epsilon\alpha^2\beta^2\delta^4$
\end{tabular}
\end{itemize}

\bigskip

These three models can also be expressed as log-linear models. We denote the state of $X$ by $i$, the state of $Y$ by $j$, and the state of $D$ by $k$. If $n_{ijk}$ describes the expected cell counts in a $3\times 3\times 2$ contingency table, then the three models can be expressed in the following way:

\bigskip

\noindent \begin{tabular}{llcl}
\textbf{Control model:} & $\log(n_{ijk})$&$=$&$\gamma + \gamma_i^X+\gamma_j^Y+\gamma_{ij}^{XY}+k\log(\epsilon)$\\ \\
\textbf{Additive model:} & $ \log(n_{ijk})$&$=$&$\gamma + \gamma_i^X+\gamma_j^Y+\gamma_{ij}^{XY}+k\log(\epsilon)+ik\log{\alpha}$ 
\\&&& $+jk\log{\beta}$\\ \\
\textbf{Multiplicative model:} & $ \log(n_{ijk})$&$=$&$\gamma + \gamma_i^X+\gamma_j^Y+\gamma_{ij}^{XY}+k\log(\epsilon)+ik\log{\alpha}$ 
\\&&& $+jk\log{\beta} +ijk\log{\delta}$
\end{tabular}

\bigskip

\begin{figure}[!t]
\centering
\includegraphics[scale=0.7]{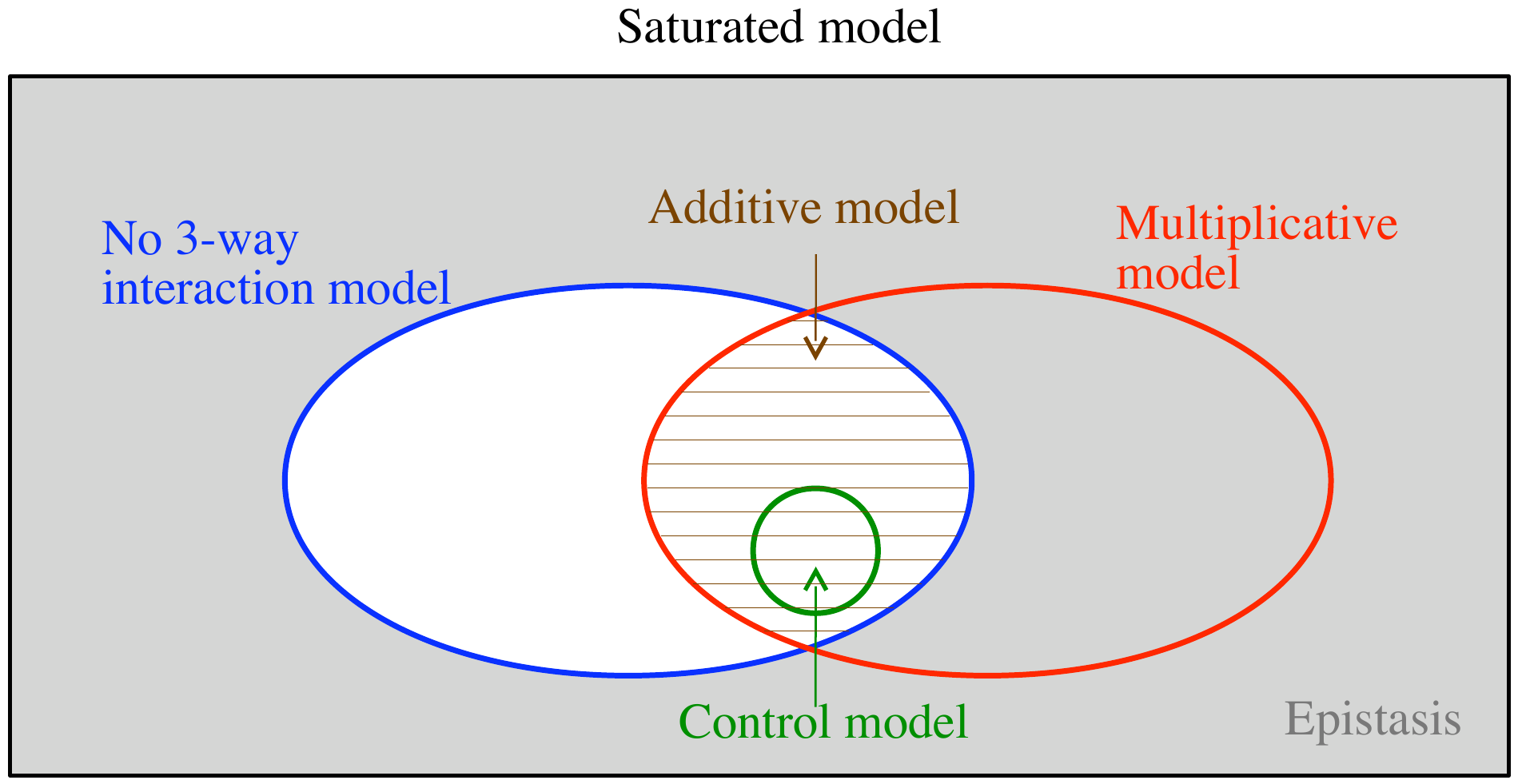}
\caption{Nesting relationship of the control model, the additive model, and the multiplicative model. The intersection of the no 3-way interaction model with the multiplicative model corresponds to the additive model. The shading indicates the presence of epistasis.}
\label{subsets}
\end{figure}

\noindent This representation explains the nesting relationship shown on the Venn diagram in Figure \ref{subsets}. Note that the additive model corresponds to the intersection of the no 3-way interaction model with the multiplicative model, and the control model is nested within the additive model.

In a biological context, interaction between markers (or SNPs) is in general used as synonym for \emph{epistasis}. \cite{Cordell} gives a broad definition: ``Epistasis refers to departure from `independence' of the effects of different genetic loci in the way they combine to cause disease''. Epistasis is for example the result of a multiplicative effect between two markers (i.e.~$\delta \neq 1$ in the multiplicative model).

In contrast, in a mathematical context interaction is used as synonym for \emph{correlation}. Two markers are said to be interacting if they are  correlated, i.e. $$\mathbb{P}(\textrm{marker 1}=i, \textrm{marker 2}=j)\neq \mathbb{P}(\textrm{marker 1}=i)\mathbb{P}(\textrm{marker 2}=j).$$

In general, in association studies the goal is to find a set of markers that are correlated with a specific phenotype. However, the markers can be correlated with each other as well. In what follows, we will use the term interaction as synonym for correlation and the term epistasis with respect to a specific phenotype synonymously to the presence of a $k$-way interaction ($k\geq 3$) between $k-1$ SNPs and a discrete phenotype.

\subsection{Algorithm}
\label{algorithm}

The $\chi^2$ goodness-of-fit-test is the most widely used test for detecting interaction within contingency tables. Under independence the $\chi^2$ statistic is asymptotically $\chi^2$ distributed. However, this approximation is problematic when the cell counts are small, which is often the case in contingency tables resulting from association studies. The other widely used test is Fisher's exact test. As its name suggests, it has the advantage of being exact. But it is a permutation test and therefore  computationally more intensive. For tables with large total counts or tables of higher dimension enumerating all possible tables with the given margins is not feasible. 

\cite{Diaconis} describe an extended version of Fisher's exact test using Markov bases. This test can be used for analyzing multidimensional tables with large total counts and the resulting posterior distribution is a good approximation of the exact distribution of the $\chi^2$-statistic even if some cell counts are small. Useful properties of Markov bases can be found in (\cite{Oberwolfach}).

The Markov basis of the null model can be computed using the software \verb+4ti2+\footnote{http://www.4ti2.de/}. Then an MCMC chain is started in the observed $3\times 3\times 2$ data table using the elements of the Markov basis as moves in the Metropolis-Hastings steps. At each step the $\chi^2$ statistic is computed. Its posterior distribution is an approximation of the exact distribution of the $\chi^2$ statistic. 



\subsubsection{Interaction tests with the extended version of Fisher's exact test}
\label{interaction_tests}

In this subsection we present various hypotheses that can easily be tested using Markov bases and discuss a hypothesis that is particularly interesting for association studies. The corresponding Markov basis can be found in the appendix. For simplicity we again constrain this discussion to the case of two SNPs and a binary phenotype.

Table \ref{loglin} consists of the standard log-linear models on three variables. Their fit to a given data table can be computed using the extended version of Fisher's exact test. We use the notation presented in (\cite{Bishop}) to denote the different models. Interaction is assumed between the variables not separated by commas in the model. So the model $(X,Y,D)$ in Table \ref{loglin} represents the independence model, the model $(XY, XD, YD)$ the no 3-way interaction model and the other models are intermediate models. For association studies the no 3-way interaction model $(XY,XD,YD)$ is particularly interesting and will be used as null model in our testing procedure.

\begin{table}[!t]
\caption{Standard interaction models for three-dimensional contingency tables.} \label{loglin} \centering \bigskip
\begin{tabular}{|c|c|c|}
\hline \textbf{Model} & \textbf{Minimal sufficient statistics} & \textbf{Expected counts}\\ \hline $(X,Y,D)$ & $(n_{i..})$, $(n_{.j.})$, $(n_{..k})$ & $\hat{n}_{ijk}=\frac{n_{i..}n_{.j.}n_{..k}}{(n_{...})^2}$\\ $(XY,D)$ & $(n_{ij.})$, $(n_{..k})$ & $\hat{n}_{ijk}=\frac{n_{ij.}n_{..k}}{(n_{...})}$ \\$(XD,Y)$ & $(n_{i.k})$, $(n_{.j.})$ & $\hat{n}_{ijk}=\frac{n_{i.k}n_{.j.}}{(n_{...})}$\\ $(X,YD)$ & $(n_{i..})$, $(n_{.jk})$ & $\hat{n}_{ijk}=\frac{n_{.jk}n_{i..}}{(n_{...})}$\\ $(XY,YD)$ & $(n_{ij.})$, $(n_{.jk})$ & $\hat{n}_{ijk}=\frac{n_{ij.}n_{.jk}}{(n_{.j.})}$\\ $(XY,XD)$ & $(n_{ij.})$, $(n_{i.k})$ & $\hat{n}_{ijk}=\frac{n_{ij.}n_{i.k}}{(n_{i..})}$\\ $(XD,YD)$ & $(n_{i.k})$, $(n_{.jk})$ & $\hat{n}_{ijk}=\frac{n_{i.k}n_{j.k}}{(n_{..k})}$\\ $(XY, XD, YD)$ & $(n_{ij.})$, $(n_{i.k})$, $(n_{.jk})$ & Iterative proportional fitting\\ \hline
\end{tabular}
\end{table}

Performing the extended version of Fisher's exact test involves sampling from the space of contingency tables with fixed minimal sufficient statistics and computing the $\chi ^2$ statistic. So the minimal sufficient statistics and the expected counts for each cell of the table need to be calculated. These are given in Table \ref{loglin}. In (\cite{Bishop}) it is shown that the cell counts cannot directly be estimated when a closed loop is present in the model configuration as for example in the no 3-way interaction model (i.e. this model can be rewritten as $(XY, YD, DX)$). But in this case, estimates can be achieved by iterative proportional fitting.


It is important to note that testing for epistasis implies working with multidimensional contingency tables and is not possible in the collapsed two-dimensional haplotype table shown below. The sufficient statistics for the model described in Table \ref{haplo} are the row and column sums $(n_{ij.})$ and $(n_{..k})$. So testing for association in this collapsed table is the same as using $(XY,D)$ as null model. In this case, the null hypothesis would be rejected even in the presence of marginal effects only, showing that testing for epistasis in Table \ref{haplo} is impossible. 

\begin{table}[!b]
\caption{Testing for association between haplotypes and phenotype.} \label{haplo} \centering \bigskip
\begin{tabular}{c c |c c|c}
\quad &\quad & \multicolumn{2}{|c|}{\textbf{Phenotype status:}} & \quad \textbf{Total:}\\ & & 0 \qquad & 1 \qquad &  \quad
\\ \hline
\, \textbf{Haplotype:} & \, 00 \quad & \qquad $n_{000}$ \qquad \qquad & \qquad $n_{001}$ \qquad \qquad & \qquad $n_{00.}$ \qquad
\\ & \, 01 \quad & \qquad $n_{010}$ \qquad \qquad & \qquad $n_{011}$ \qquad \qquad & \qquad $n_{01.}$ \qquad \\ & \, 10 \quad & \qquad $n_{100}$ \qquad \qquad & \qquad $n_{101}$ \qquad \qquad & \qquad $n_{10.}$ \qquad
\\ & \, 11 \quad & \qquad $n_{110}$ \qquad \qquad & \qquad $n_{111}$ \qquad \qquad & \qquad $n_{11.}$ \qquad \\ \hline
\textbf{Total:} &  & \qquad $n_{..0}$  \qquad \qquad & \qquad $n_{..1}$ \qquad \qquad &
\qquad $n_{...}$ \qquad
\end{tabular}
\end{table}

\subsubsection{Hypothesis testing with the extended version of Fisher's exact test}

Our goal is to detect epistasis when present. According to our definition of epistasis in Section \ref{interaction_models} and as shown in Figure \ref{subsets}, epistasis is present with regard to two SNPs and a specific phenotype, when a 3-way interaction is found. So we suggest using as null hypothesis the no 3-way interaction model and testing this hypothesis with the extended version of Fisher's exact test. The corresponding Markov basis consists of 15 moves and is given in the appendix. It can be used  to compute the posterior distribution of the $\chi^2$ statistic and approximate the exact p-value of the data table. If the p-value is lower than some threshold, we reject the null hypothesis of no epistasis.

Although in this paper we focus merely on epistasis, it is worth noting that one can easily build tests for different types of interaction using Markov bases. If one is interested in detecting whether the epistatic effect is of multiplicative nature, one can perform the extended version of Fisher's exact test on the contingency tables, which have been classified as epistatic, using the multiplicative model as null hypothesis. In this case, the corresponding Markov basis consists of 49 moves. Similarly, if one is interested in detecting additive effects, one can use the additive model as null hypothesis and test the contingency tables, which have been classified as non-epistatic. In this case, the corresponding Markov basis consists of 156 moves. The Markov bases for these tests can be found on our website\footnote{http://www.carolineuhler.com/epistasis.htm}.

\section{Results}
\label{results}

In this section, we first conduct a simulation study to evaluate the performance of the suggested method. We then compare our method to a two-stage logistic regression approach and BEAM (\cite{BEAM}). Logistic regression is a widely used method for detecting epistasis within a selection of SNPs. BEAM is a purely Bayesian method for detecting epistatic interactions on a genome-wide scale. We end this section by applying our method to a genome-wide data set consisting of 685 dogs with the goal of finding epistasis associated with canine hair length. 

\subsection{Simulation study}
\label{simulation}

We simulated a total of 50 potential association studies with 400 cases and 400 controls for three different minor allele frequencies of the causative SNPs and the three models of interaction presented in Section \ref{interaction_models}. We chose as minor allele frequencies (MAF) 0.1, 0.25 and 0.4. The parameters for the three models of interaction were determined numerically fixing the effect size 
$$\lambda_i:=\frac{p(D=1|g_i=1)}{p(D=0|g_i=1)}\frac{p(D=0|g_i=0)}{p(D=1|g_i=0)}-1$$
and the prevalence 
$$\pi:=\sum_{g_1,g_2}p(D|g_{1},g_{2})p(g_{1},g_{2}).$$
For our simulations, we used an effect size of $\lambda_1=\lambda_2=1$ and a sample prevalence of $\pi=0.5$. Choosing in addition $\alpha=\beta$ in the additive model, and $\alpha=\beta$ and $\delta=3\alpha$ in the multiplicative model determines all parameters of the interaction models and one can solve for $\alpha, \beta, \delta$ and $\epsilon$ numerically.

The simulations were performed using HAP-SAMPLE (\cite{HAP-SAMPLE}) and
were restricted to the SNPs typed with the Affy CHIP on chromosome 9 and chromosome
13 of the Phase I/II HapMap data\footnote{http://hapmap.ncbi.nlm.nih.gov/}, resulting in about 10,000 SNPs per individual. On each of the two chromosomes we selected
one SNP to be causative. The causative SNPs were chosen consistent with the minor allele frequencies and far apart from any other marker (at least 20,000bp apart) in order to avoid linkage with nearby SNPs. Note that HAP-SAMPLE generates the cases and controls by resampling from HapMap. This means that the simulated data shows linkage disequilibrium and allele frequencies similar to real data.

As suggested by \cite{Marchini}, we took a two-stage approach for finding interacting SNPs. In the first step, we ranked all SNPs according to their p-value in Fisher's exact test and selected the ten SNPs with the lowest marginal p-values. Within this subset, we then tested for interaction using the extended version of Fisher's exact test and the no 3-way interaction model as null hypothesis. We generated three Markov chains with 40,000 iterations each and different starting values, and used the tools described by Toft et al. (2007) and Gilks, Richardson, and Spiegelhalter (1995) to assess convergence of the chains. This included analyzing the Gelman-Rubin statistic and the autocorrelations. After discarding an initial burn-in of 10,000 iterations, we combined the remaining samples of the three chains to generate the posterior distribution of the $\chi ^2$ statistic.

\begin{figure}[!h]
\centering
\includegraphics[scale=0.19]{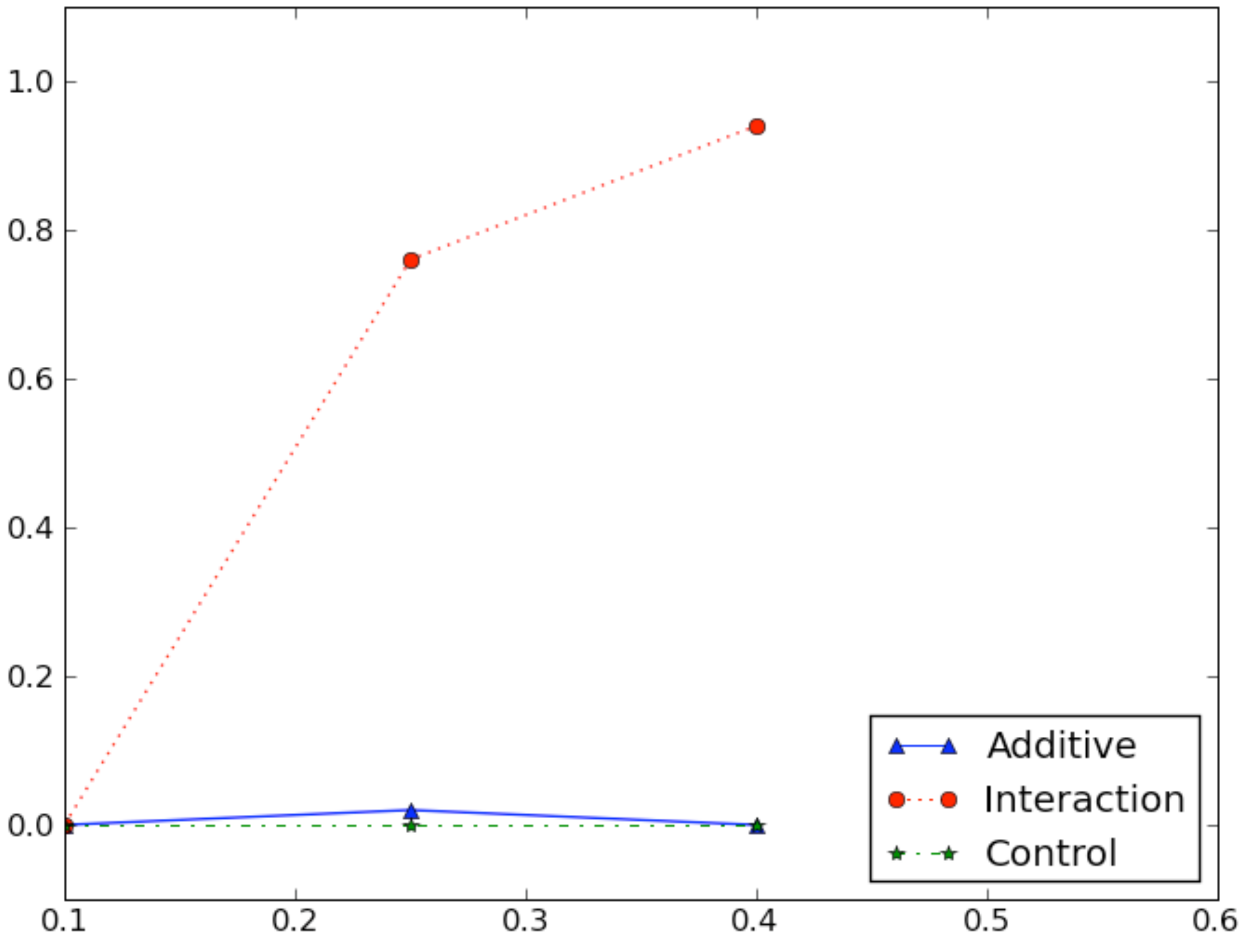}
\includegraphics[scale=0.19]{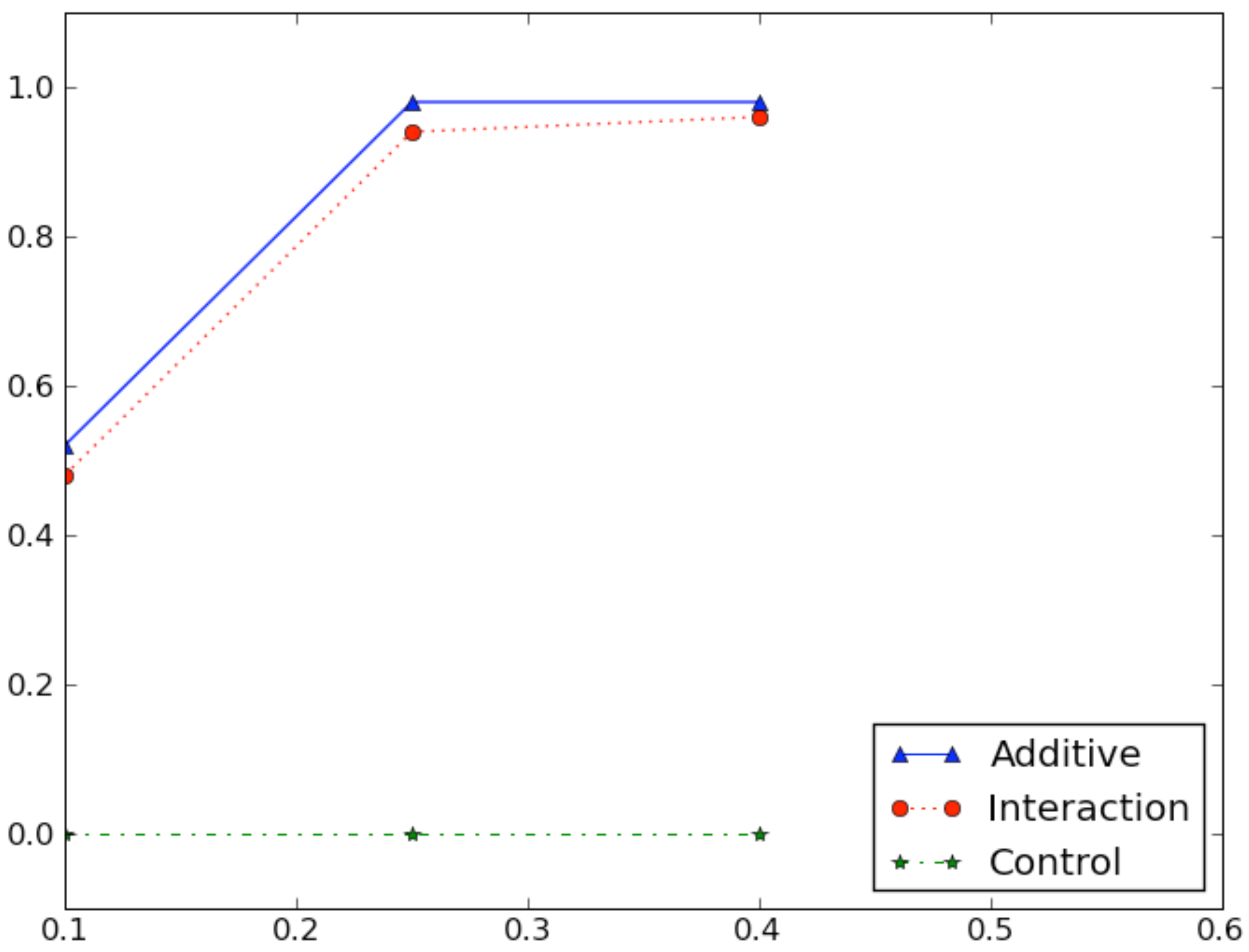}
\includegraphics[scale=0.19]{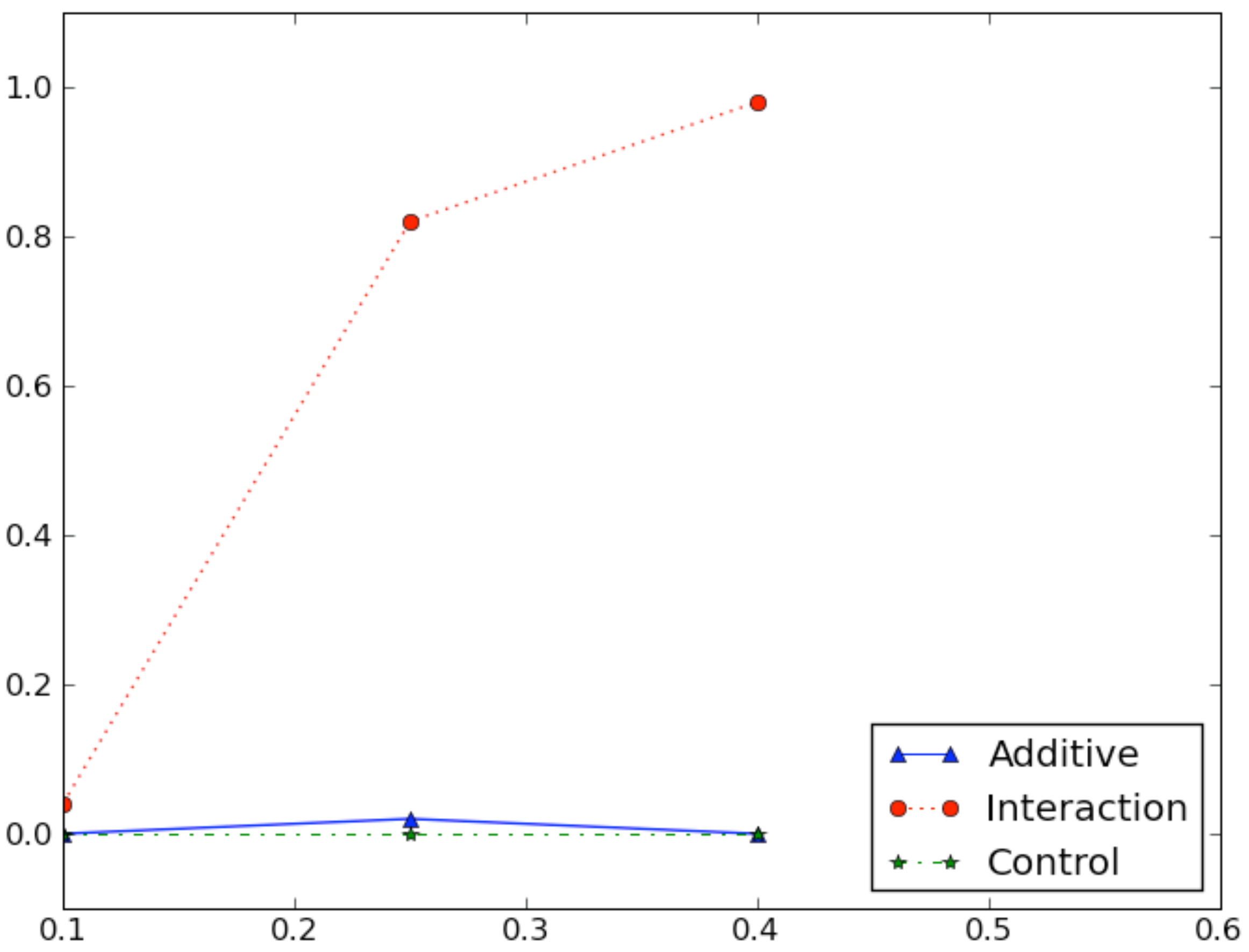}
\caption{Rejection rate of the no 3-way interaction test in the two-stage approach on 50 simulated association studies for MAF=0.1, MAF=0.25, and MAF=0.4 (left). Proportion of 50 association studies, in which the two causative SNPs were ranked among the ten SNPs with the lowest p-values by Fisher's exact test (middle). Rejection rate of the no 3-way interaction hypothesis using only the extended version of Fisher's exact test on the 50 causative SNP pairs (right).}
\label{power}
\end{figure}

In Figure \ref{power} (left), we report the rejection rate of the no 3-way interaction hypothesis for each of the three minor allele frequencies. Per point in the figure we simulated 50 potential association studies. The power of our two-stage testing procedure corresponds to the curve under the multiplicative model. The higher the minor allele frequency, the more accurately we can detect epistasis. Under the additive model and the control model, no epistasis is present. We never rejected the null hypothesis under the control model and only once under the additive model, resulting in a high specificity of the testing procedure. 

We also analyze the performance of each step separately. Figure \ref{power} (middle) shows the performance of the first step and reports the proportion of 50 association studies, in which the two causative SNPs were ranked among the ten SNPs with the lowest p-values. Because Fisher's exact test measures marginal association, the curves under the additive model and the multiplicative model are similar. 

Figure \ref{power} (right) shows the performance of the second step in our method and reports the proportion of 50 association studies, in which the null hypothesis of no 3-way interaction was rejected using only the extended version of Fisher's exact test on the 50 causative SNP pairs.

\subsection{Comparison to logistic regression}
\label{logistic_regression}

For validation, we compare the performance of our method to logistic regression via ROC curves. Logistic regression is probably the most widely used method for detecting epistasis within a selection of SNPs nowadays. We base the comparison on the simulated association studies presented in the previous section using only the simulations under the multiplicative model. The structure of interaction within this model should favor logistic regression as logistic regression tests for exactly this kind of interaction.

As before, for each minor allele frequency and each of the 50 simulation studies we first filtered all SNPs with Fisher's exact test and chose the ten SNPs with the lowest p-values for further analysis. The causative SNPs are within the ten filtered SNPs for 19 (46) [45] out of the 50 simulation studies for MAF=0.1 (MAF=0.25) [MAF=0.4]. We then ran the extended version of Fisher's exact test and logistic regression on all possible pairs of SNPs in the subsets consisting of the ten filtered SNPs. This results in $50 \cdot {10\choose 2}$ tests per minor allele frequency with 19 (46) [45] true positives for MAF=0.1 (MAF=0.25) [MAF=0.4].

\begin{figure}[!b]
\centering
\includegraphics[scale=0.27]{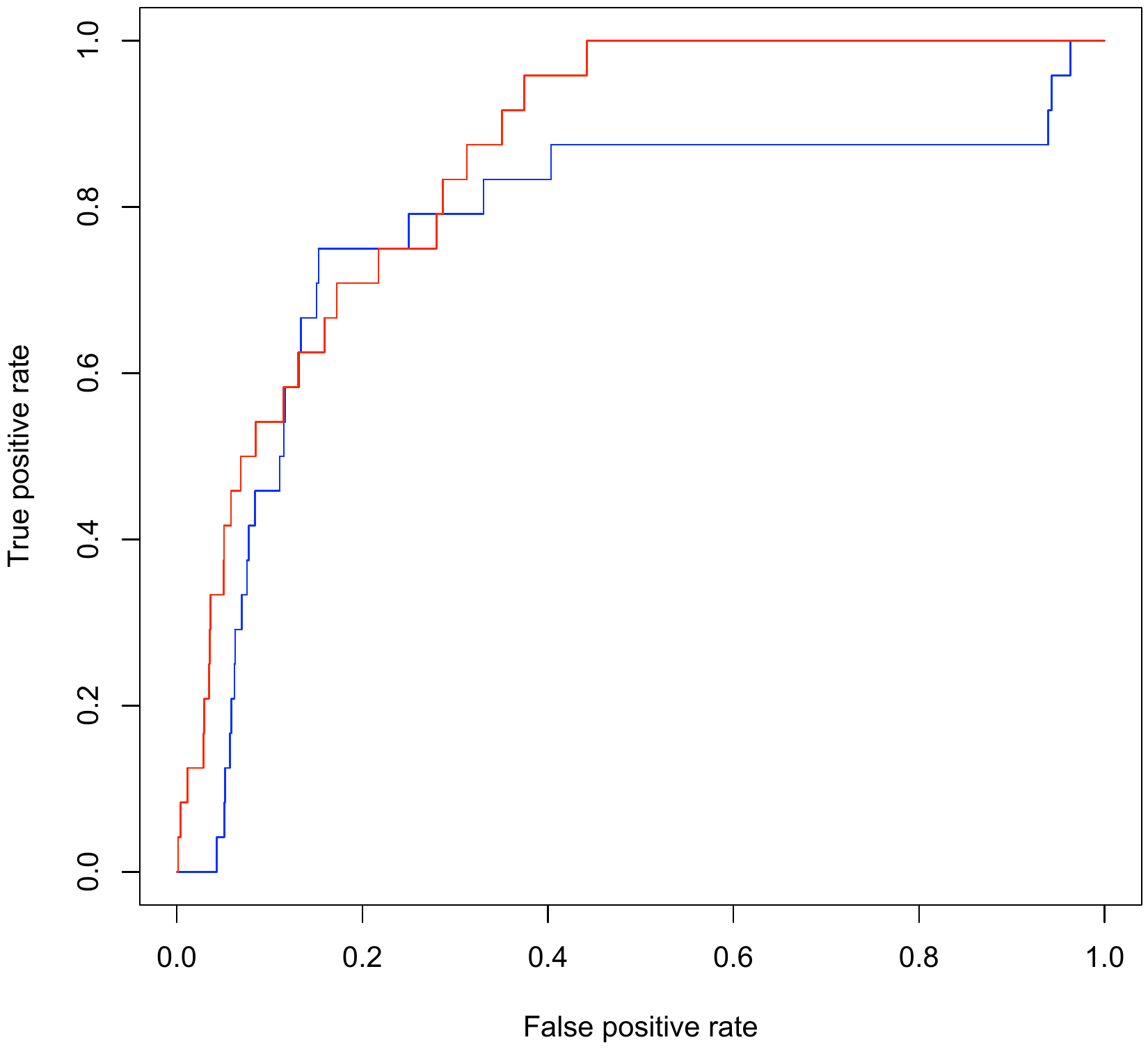}
\includegraphics[scale=0.27]{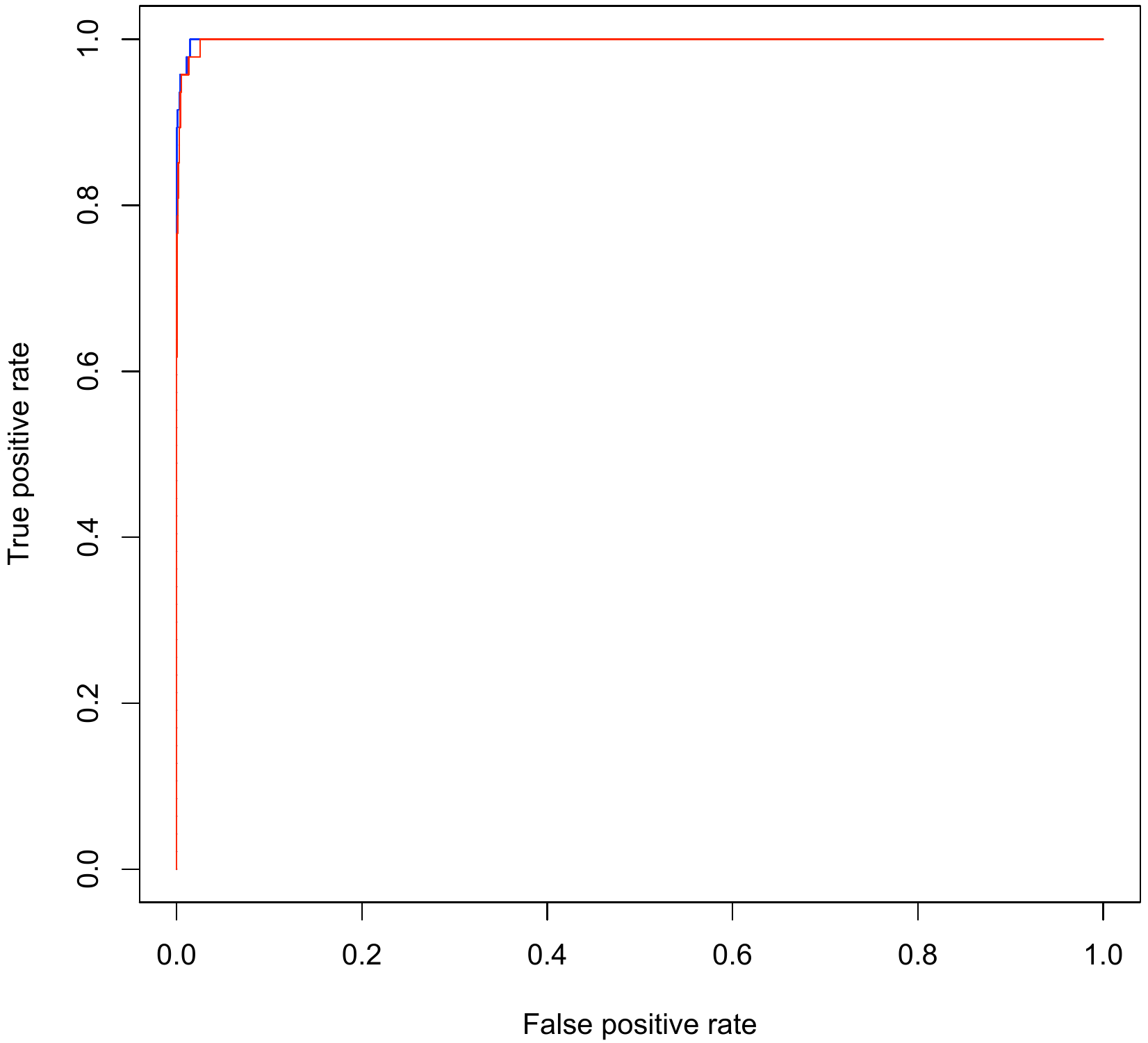}
\includegraphics[scale=0.27]{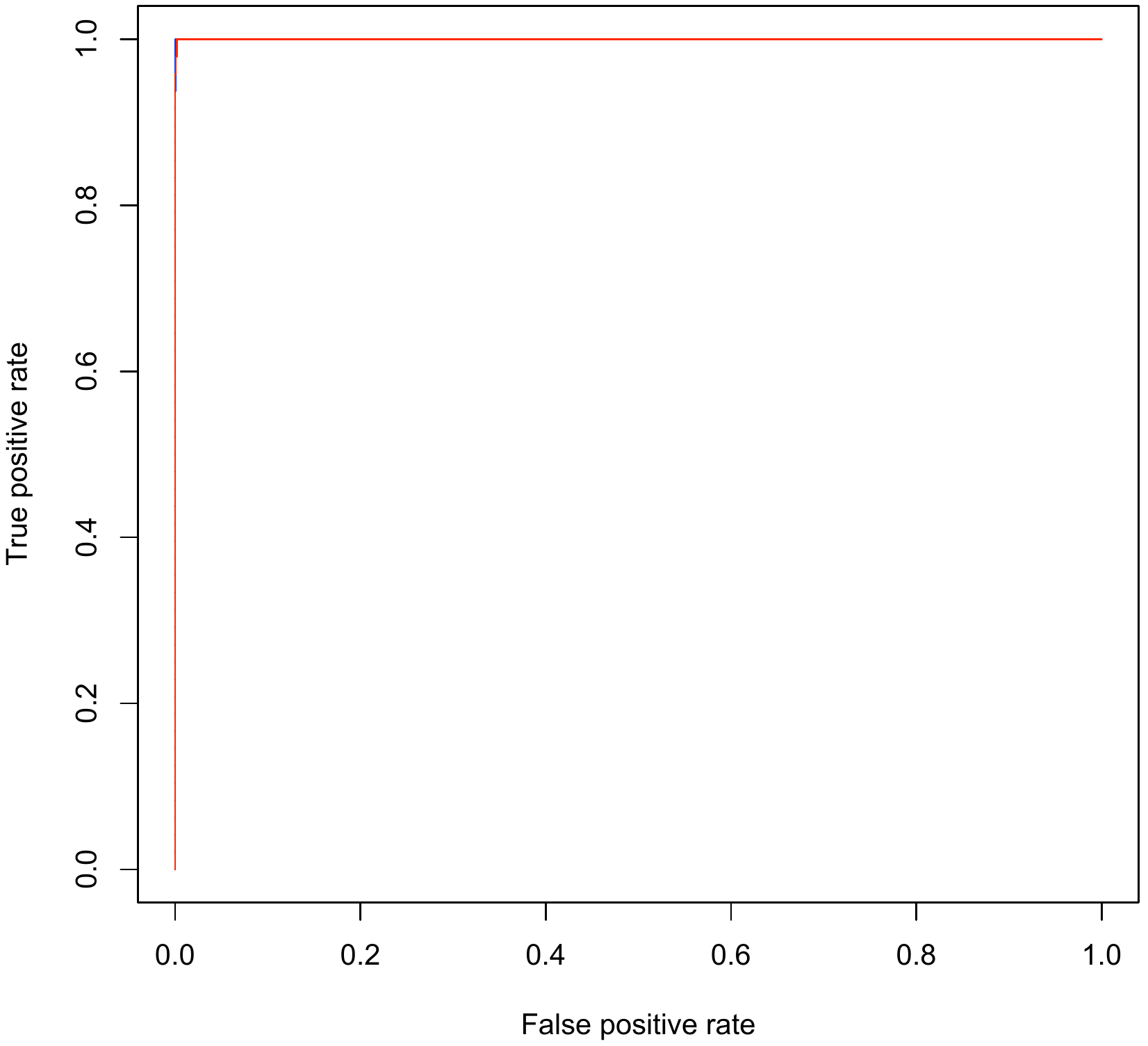}
\caption{ROC curves of the extended version of Fisher's exact test (in red) and of logistic regression (in blue) for MAF=0.1 (left), MAF=0.25 (middle), and MAF=0.4 (right).}
\label{ROC}
\end{figure}

The ROC curves comparing the second stage of our method to a logistic regression approach are plotted in Figure \ref{ROC} showing that our method does significantly better than logistic regression for MAF=0.1 with an area under the ROC curve of 0.861 compared to 0.773 for logistic regression. For MAF=0.25 and MAF=0.4 both methods have nearly perfect ROC curves with areas 0.9986 [0.99994] for our method compared to 0.9993 [0.99997] for logistic regression for MAF=0.25 [MAF=0.4].

\subsection{Comparison to BEAM}
\label{BEAM_simulation}

We also compare our method to BEAM, a Bayesian approach for detecting epistatic interactions in association studies (\cite{BEAM}). We chose to compare our method to BEAM, because the authors show it is significantly more powerful than a variety of other approaches including the stepwise logistic regression approach, and it is one of the few recent methods that can handle genome-wide data. 

In this method, all SNPs are divided into three groups, namely, SNPs that are not associated with the disease, SNPs that contribute to the disease risk only through main effects, and SNPs that interact to cause the disease. BEAM outputs the posterior probabilities for each SNP to belong to these three groups. The authors propose to use the results in a frequentist hypothesis-testing framework calculating the so called B-statistc and testing for association between each SNP or set of SNPs and the disease phenotype. BEAM was designed to increase the power to detect any association with the disease, and not to separate main effects from epistasis. Therefore, BEAM outputs SNPs that interact marginally OR through a k-way interaction with the disease. This does not match our definition of epistasis since using BEAM the presence of marginal effects only already gives rise to a significant result. 

We compare our method to BEAM using the B-statistic. BEAM reports this statistic only for the pairs of SNPs which have a non-zero posterior probability of belonging to the third group. In addition, the B-statistic is automatically set to zero for the SNP pairs which are found to be interacting marginally with the disease. We force BEAM to include the marginal effects into the B-statistic by choosing a significance level of zero for marginal effects. This should be favorable for BEAM in terms of sensitivity.

We ran BEAM with the default parameters on our simulated datasets for the multiplicative model. Due to the long running time of BEAM, we based the comparison only on 1,000 SNPs out of the 10,000 SNPs simulated for the analysis in Section \ref{simulation}. BEAM takes about 6.8 hours for the analysis of one dataset with 10,000 SNPs and 400 cases and controls, whereas the same analysis with our method takes about 0.8 hours on an Intel Core 2.2 GHz laptop with 2 Gb memory.

\begin{figure}[!t]
\centering
\includegraphics[scale=0.27]{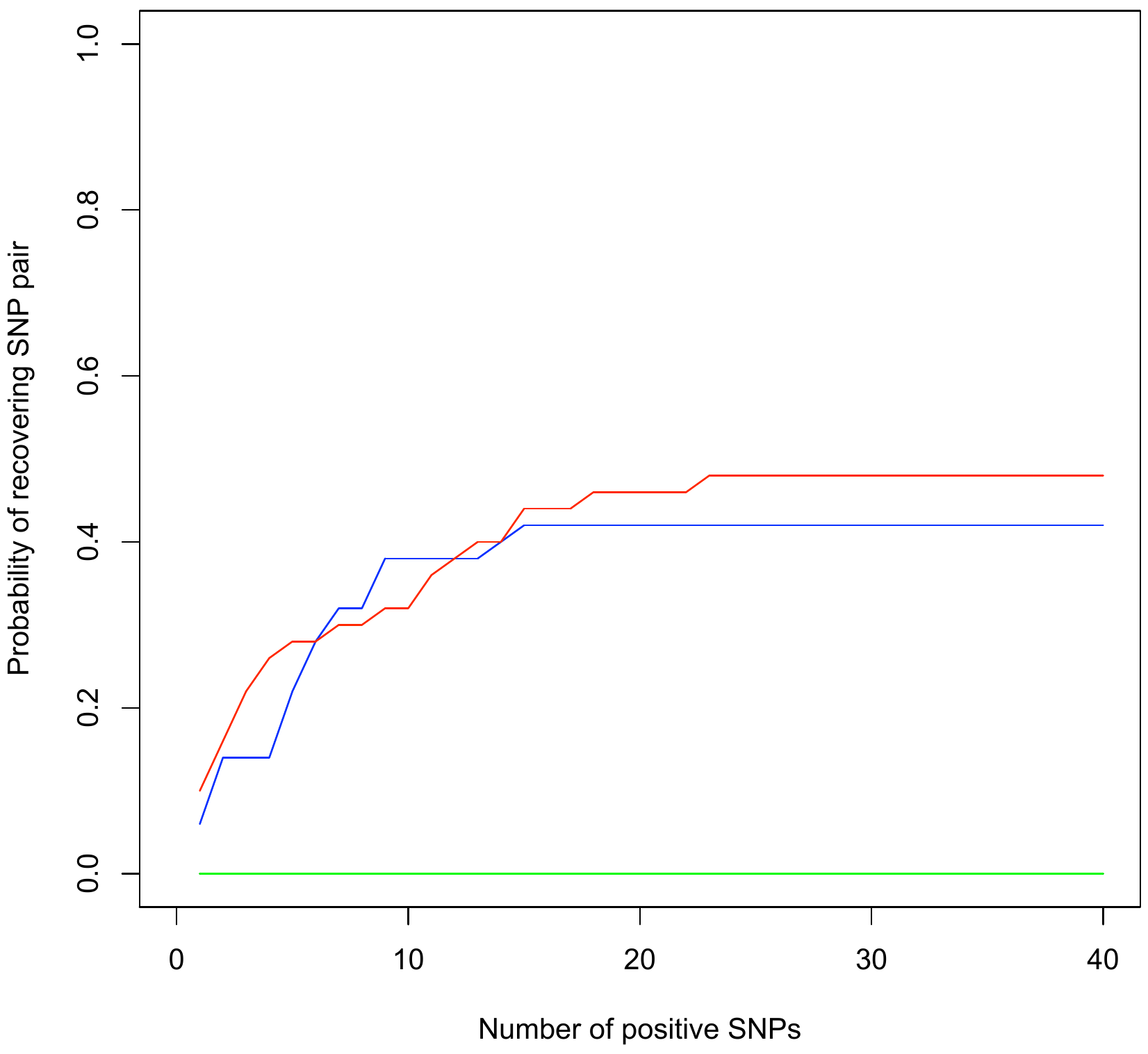}
\includegraphics[scale=0.27]{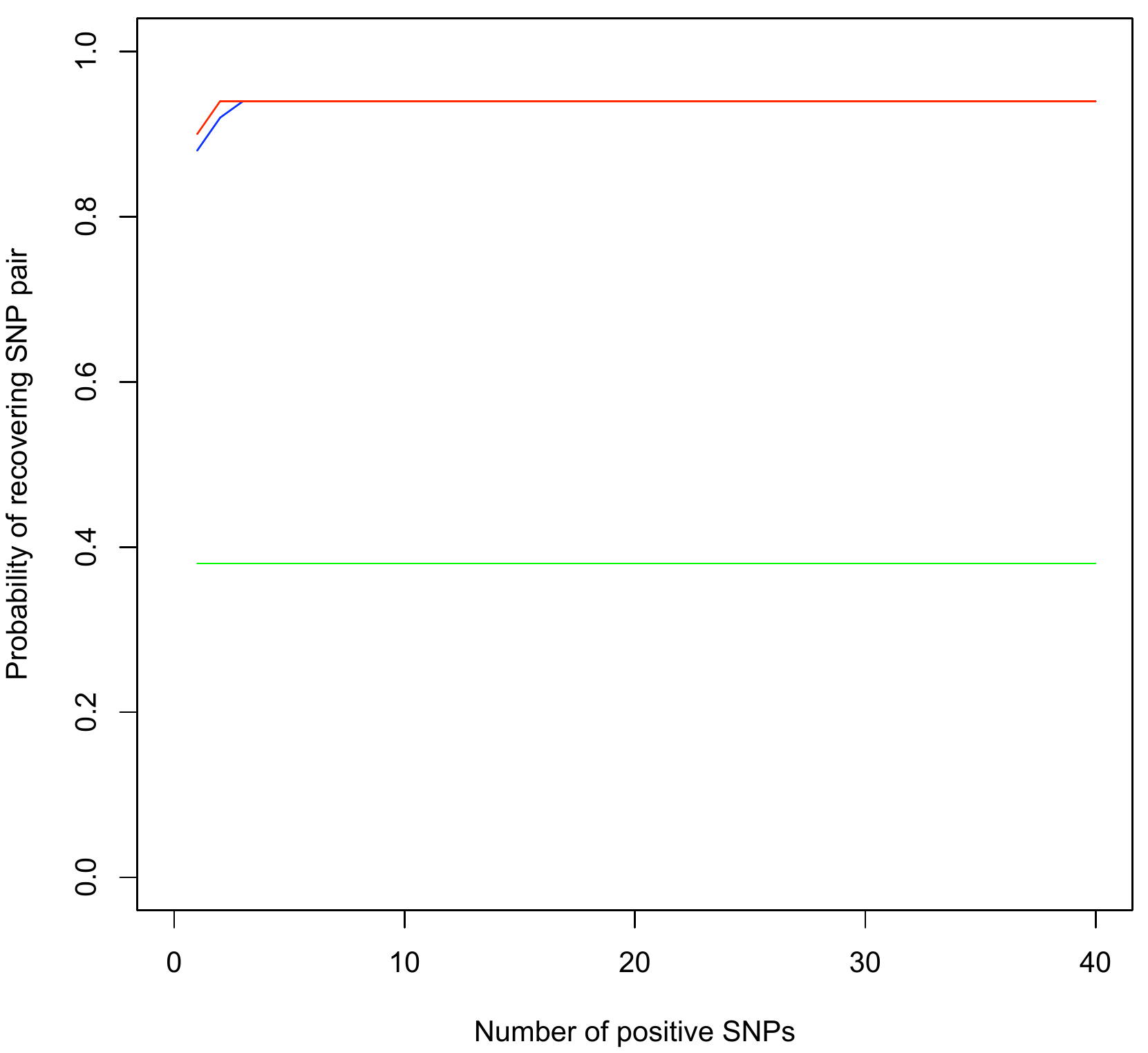}
\includegraphics[scale=0.27]{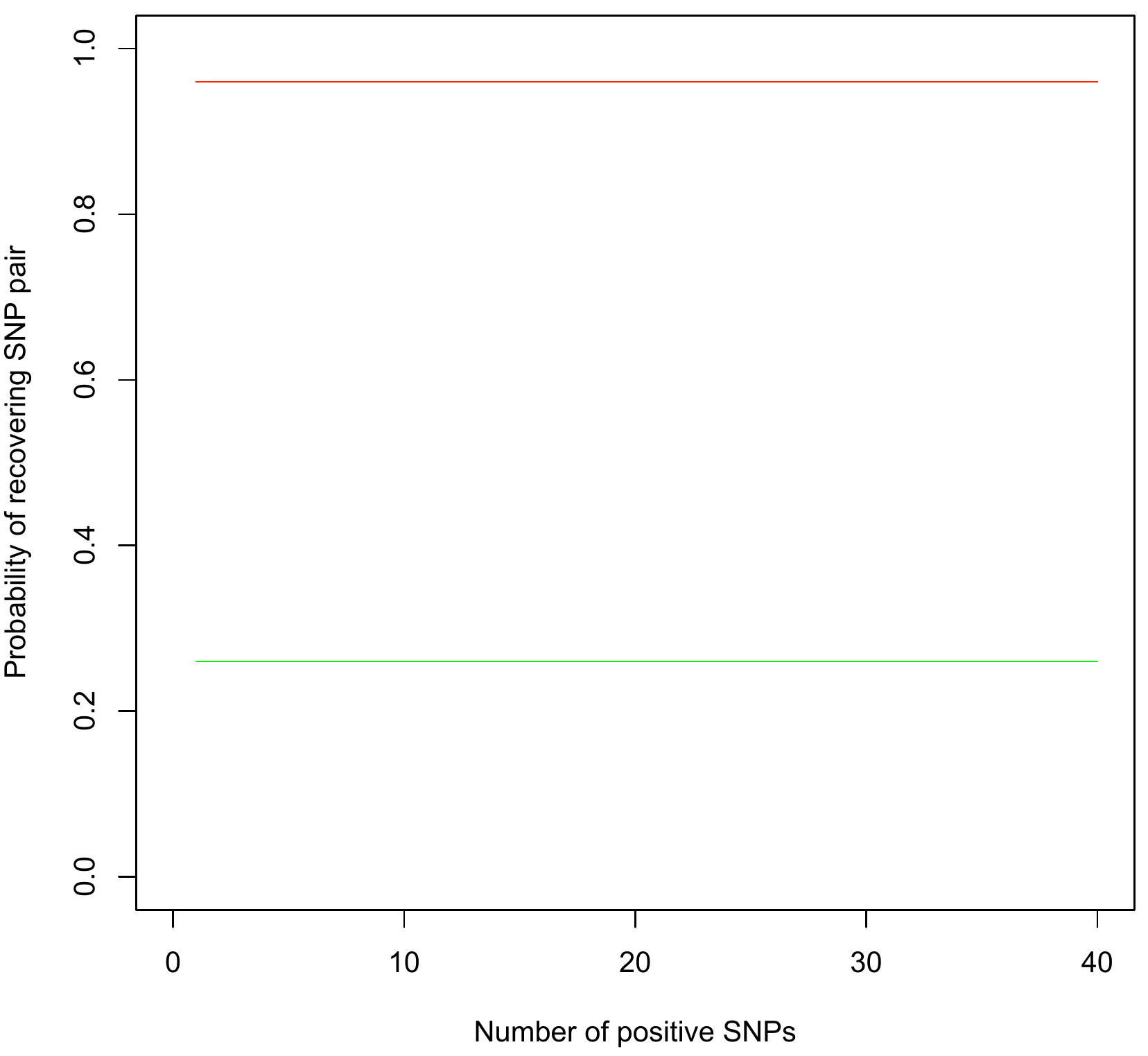}
\caption{Proportion of simulation studies for which the interacting SNP pair belongs to the $x$ SNPs with the lowest p-values for MAF=0.1 (left), MAF=0.25 (middle), and MAF=0.4 (right). The performance of our method is shown in red, stepwise logistic regression in blue, and BEAM in green.}
\label{ROC_BEAM}
\end{figure}

In contrast to BEAM, our method is a stepwise approach, which makes a comparison via ROC curves difficult. We therefore compare the performance of all three tests by plotting for a fixed number $x$ of SNPs the proportion of simulation studies for which the interacting SNP pair belongs to the $x$ SNPs with the lowest p-values. The resulting curves are shown in Figure \ref{ROC_BEAM}. Although the marginal effects were not extracted, BEAM has a very high false negative rate, attributing a p-value of 1 to the majority of SNPs, interacting and not interacting ones.

\subsection{Genome-wide association study of hair length in dogs}
\label{dogs}

We demonstrate the potential of our Markov basis method in genome-wide
association studies by analyzing a hair length dataset consisting of $685$ dogs from $65$ breeds and containing $40'842$ SNPs (\cite{CNQ+09}).

The individuals in (\cite{CNQ+09}) were divided into two groups for the hair
length phenotype: $319$ dogs from $31$ breeds with long hair as cases
and $364$ from $34$ breeds with short hair as controls. In the original
study, it is shown that the long versus short hair phenotype is associated
with a mutation (Cys95Phe) that changes exon one in the \emph{fibroblast
growth factor-5} (\emph{FGF5} gene). Indeed, the SNP with the lowest
p-value using Fisher's exact test 
is located on chromosome 32 at position $7,100,913$ for the
Canmap dataset, i.e.~about $300$Kb apart from \emph{FGF5}. 

We ranked the $40,842$ SNPs by their p-value using Fisher's exact test
and selected the 20 lowest ranked SNPs (about $5\%$) to test for 3-way interaction. Note that all 20 SNPs are significantly correlated
with the phenotype. We found a significant p-value for four out of the 
${20\choose 2}$ pairs. These pairs together with their p-values are listed in Table
\ref{tab:SNPs-significantly-interacting}.

\begin{table}
\begin{centering}
\begin{tabular}{|c|c|c|}
\hline 
chromosome and location of SNPs & p-value & potential relevant genes \tabularnewline
\hline 
chr30.18465869, chr26.6171079 & 0 & \emph{FGF7}-?\tabularnewline
\hline 
chr15.44092912, chr23.49871523 & 0 & \emph{IGF1}-\emph{P2RY1}\tabularnewline
\hline 
chr24.26359293, chr15.43667654 & 2e-04 & \emph{ASIP}-?\tabularnewline
\hline 
chr15.43667654, chr23.49871523 & 1e-04 & ?-\emph{P2RY1}\tabularnewline
\hline
\end{tabular}
\par\end{centering}

\caption{\label{tab:SNPs-significantly-interacting}Pairs of SNPs, which significantly
interact with the hair length phenotype for the Canmap dataset. }

\end{table}

The pairs include six distinct SNPs located on five different chromosomes and the two SNPs lying on the same chromosome are not significantly interacting (p-value of $0.54$). This means that a false positive correlation due to hitchhiking effects can be avoided. Hitchhiking effects are known to extend across long stretches of chromosomes in particular in domesticated species (\cite{MCP+07}) consistent with the prediction of \cite{MSJ74}.

In order to identify potential pathways we first considered genes, which are
close to the six SNPs we identified as interacting. To do so, we used
the dog genome available through the ncbi website\footnote{http://www.ncbi.nlm.nih.gov/genome/guide/dog/}.
Most of the genes we report here have been annotated automatically.
Our strategy was to consider the gene containing the candidate SNP (if any)
and the immediate left and right neighboring gene, resulting in a total of two or three genes per SNP. 

Among the six significantly interacting SNPs, four are located close
to genes that have been shown to be linked to hair
growth in other organisms. This is not surprising, since these SNPs are significantly marginally
associated with hair growth. We report
here the function of these candidate genes. The two other SNPs are located close to genes that we were not able to identify as functionally related to hair growth. 

The SNP chr30.18465869 is located close to (about $80$Kb) \emph{fibroblast
growth factor 7} (\emph{FGF7} also called \emph{keratinocyte growth
factor}, \emph{KGF}), i.e.~it belongs to the same family as the gene
reported in the original study (but on a different chromosome). The FGF
family members are involved in a variety of biological processes including
hair development reported in human, mouse, rat and chicken (GO:0031069, \cite{ABB+00}).

Secondly, chr15.44092912 is located between two genes, and about
$200$Kb from the \emph{insulin-like growth factor 1} gene (\emph{IGF1}).
\emph{IGF1} has been reported to be associated with the hair growth
cycle and the differentiation of the hair shaft in mice (\cite{WS05}). 

Thirdly, chr23.49871523 is located about $430$Kb from the \emph{purinergic receptor P2Y1} (\emph{P2RY1}). The purinergic receptors have been shown to be part of a signaling system for proliferation
and differentiation in human anagen hair follicles (\cite{GLB08}). 

Finally, chr24.26359293 is located inside the agouti-signaling protein
(gene \emph{ASIP}), a gene known to affect coat color in dogs and other 
mammals. The link to hair growth is not obvious but this gene is expressed during 4-7 days of hair growth in mice (\cite{WSF+07}).

According to our analysis, \emph{IGF1} and \emph{P2RY1}
are significantly interacting. All other pairs of interacting SNPs involve at least one SNP for which we were not able to identify a closeby candidate gene, which is related to hair growth (see Table
\ref{tab:SNPs-significantly-interacting}). \emph{IGF1} has a tyrosine
kinase receptor and \emph{P2RY1} is a G-protein coupled receptor.
One possibility is that these receptors cross-talk 
as has been shown previously for these types of receptors in order to control
mitogenic signals (\cite{DB99}). To
conclude, a functional assay would be necessary to establish that
any of the statistical interactions we found are also biologically meaningful.

We also considered all triplets of SNPs among the 20 preselected SNPs
and tested for 4-way interaction. However, we did not find any evidence for
interaction among the ${20\choose 3}$ triplets.

\section{Discussion}
In this paper, we proposed a Markov basis approach for detecting epistasis in genome-wide association studies. The use of different Markov bases allows to easily test for different types of interaction and epistasis involving two or more SNPs. These Markov bases need to be computed only once and can be downloaded from our website\footnote{http://www.carolineuhler.com/epistasis.htm} for the tests presented in this paper. 


We tested our method in simulation studies and showed that it outperforms a stepwise logistic regression approach and BEAM for the multiplicative interaction model. Logistic regression has the advantage of a very short running time (3 seconds compared to 32 minutes for the analysis of one dataset with 10,000 SNPs and 400 cases and controls not including the filtering step, which takes about 16 minutes for both methods on an Intel Core 2.2 GHz laptop with 2 Gb memory). However, especially for a minor allele frequency of 0.1 logistic regression performs significantly worse than our method, even when simulating epistasis under a multiplicative model, which should favor logistic regression. BEAM on the other hand, has the advantage of not needing to filter the large number of SNPs first. However, it runs about 8 times slower than our method for our simulations and has a very high false negative rate. One possible difference between our results and what the authors of BEAM have found might be due to linkage disequilibrium in our data. As of now, BEAM handles linkage disequilibrium with a first order Markov chain, which is likely to be improved in future versions. But as of today, we conclude that this method is impractical for whole genome association studies, since linkage disequilibrium is present in most real datasets.

The limitation of our method is the need for a filtering method to reduce the number of SNPs to a small subset. Especially if the marginal association of the interacting SNPs with the disease is small, these SNPs might not be caught by the filter. However, in our simulations using Fisher's exact test as filter seems to perform well. Another possibility is to incorporate biological information to choose a subset of possibly interacting SNPs, such as existing pathways (\cite{EMH+09}).

We demonstrated the potential of the proposed two-stage method in genome-wide
association studies by analyzing a hair length dataset consisting of $685$ dogs and containing $40'842$ SNPs. In this dataset, we found a significant epistatic effect for four SNP pairs. These SNPs lie on different chromosomes, reducing the risk of a false positive correlation due to linkage effects. Finally, we would encourage establishing if these interactions are also biologically meaningful by a functional assay.



\section*{Acknowledgements}
We would like to thank Anders Albrechsten, Heather J.~Cordell, Luke W.~Miratrix, Rasmus Nielsen, Lior Pachter, Montgomery Slatkin, Yun S.~Song and Bernd Sturmfels, for many helpful discussions. We would also like to acknowledge  Yu Zhang for help with BEAM and Heidi Parker and Elaine Ostrander for providing the dog dataset.

\section*{Appendix: Markov basis}

The Markov basis corresponding to the no 3-way interaction model on a $3\times 3\times 2$ table is given below, where the tables are reported as vectors $$(n_{111}, n_{211}, n_{311}, n_{121}, n_{221}, n_{321}, n_{131}, n_{231}, n_{331}, n_{112}, n_{212}, n_{312}, n_{122}, n_{222}, n_{322}, n_{132}, n_{232}, n_{332}).$$

\vspace{0.5cm}
\noindent \begin{tabular}{lccrrrrrrrrrrrrrrrrrrc}
$f_1$&=&(0&0&0&1&0&-1&-1&0&1&0&0&0&-1&0&1&1&0&-1)\\
$f_2$&=&(0&0&0&0&1&-1&0&-1&1&0&0&0&0&-1&1&0&1&-1)\\
$f_3$&=&(1&0&-1&0&0&0&-1&0&1&-1&0&1&0&0&0&1&0&-1)\\
$f_4$&=&(0&1&-1&0&0&0&0&-1&1&0&-1&1&0&0&0&0&1&-1)\\
$f_5$&=&(0&0&0&1&-1&0&-1&1&0&0&0&0&-1&1&0&1&-1&0)\\
$f_6$&=&(1&-1&0&0&0&0&-1&1&0&-1&1&0&0&0&0&1&-1&0)\\
$f_7$&=&(1&-1&0&-1&1&0&0&0&0&-1&1&0&1&-1&0&0&0&0)\\
$f_8$&=&(1&0&-1&-1&0&1&0&0&0&-1&0&1&1&0&-1&0&0&0)\\
$f_9$&=&(0&1&-1&0&-1&1&0&0&0&0&-1&1&0&1&-1&0&0&0)\\
$f_{10}$&=&(0&1&-1&-1&0&1&1&-1&0&0&-1&1&1&0&-1&-1&1&0)\\
$f_{11}$&=&(1&0&-1&0&-1&1&-1&1&0&-1&0&1&0&1&-1&1&-1&0)\\
$f_{12}$&=&(-1&1&0&1&0&-1&0&-1&1&1&-1&0&-1&0&1&0&1&-1)\\
$f_{13}$&=&(1&-1&0&0&1&-1&-1&0&1&-1&1&0&0&-1&1&1&0&-1)\\
$f_{14}$&=&(1&0&-1&-1&1&0&0&-1&1&-1&0&1&1&-1&0&0&1&-1)\\
$f_{15}$&=&(0&1&-1&1&-1&0&-1&0&1&0&-1&1&-1&1&0&1&0&-1)
\end{tabular}

\bibliographystyle{bepress}
\bibliography{disease_paper}

\end{document}